# Characterization of spin coated Zn doped cupric oxide thin films

P. Samarasekara [1, 2], P.G.D.C.K. Karunarathna [2] H.P. Weeramuni [1] and C.A.N. Fernando [2]

[1]Department of Physics, University of Peradeniya, Peradeniya, Sri Lanka

[2]Department of Nano Science Technology, Wayamba University of Sri Lanka, Kuliyapitiya, Sri Lanka

**Abstract**

Thin films of $Zn^{2+}$ doped cupric oxide (CuO) were synthesized using spin coating technique starting from a solution with Cu and Zn. The speed of spin coating and time duration were varied to fabricate a film with required uniform thickness. Samples were subsequently annealed in air at different annealing temperatures to crystallize the phase of CuO. Doping concentration of $Zn^{2+}$ was varied up to weight percentage of 10%. Structural properties of samples were determined using X-ray diffraction technique. Particle size, dislocation density and strain at different doping concentrations were determined using XRD patterns. Optical properties were measured by means of UV/Vis spectrometer. Optical band gap of CuO could be tailored by doping a trace amount of $Zn^{2+}$. The optical band gap decreases with increase of particle size. The particle size, dislocation density, strain and optical band gap have a turning point close to the doping concentration of 6%.

## 1. Introduction:

CuO is a black color p-type semiconductor material with band gap of 1.2-1.9eV. CuO finds potential applications in gas sensors, solar energy applications, diode fabrications, lithium ion batteries, electrochemical cells, superconductors [1] photovoltaic [2], nanoelectronics [3] and spintronics [4]. Conductivity and photoconductivity of CuO thin films deposited using sol-gel spin coating have been studied as a function of layers [2]. In addition, cuprous oxide thin films have been synthesized on silicon substrate using the sol-gel spin coating method [5]. The band gap of the films presented in above manuscripts varies in the range of 2.0-2.2 eV. Effect of applied potential on the physical properties of electrodeposited CuO thin films has been investigated [6]. CuO thin films prepared by spray pyrolysis technique have been characterized by measuring structural, electrical and optical properties [7]. XRD, absorption spectroscopy and

conductivity/ resistivity analyses have been performed for these films. 350 °C has been used as the substrate temperature in their study. A Hall-effect measurement system was utilized to measure Hall-coefficient, charge carrier concentration, mobility, resistivity and conductivity of the films [7].

A pure metal has been employed as a dopant in many oxide thin films as following. Optical and electrical properties of Zn doped tin oxide thin films prepared using sole gel spin coating method have been investigated. According to that study, the sheet resistance of tin oxide films increases with the Zn doping concentration [8]. Gas sensing properties have been improved by doping Na in spin coated ZnO thin films [9]. Gas sensitivity has been measured in $CO_2$ gas. Alloying effects on the structural and optical properties of nanocrystalline copper zinc oxide thin films synthesized by spin coating have been investigated [10]. According to their study, Zn atoms occupy Cu sites in the monoclinic structure of CuO without changing the structure of CuO. Structural and optical properties of copper doped ZnO nanoparticles and thin films have been studied. Optical ban gap of ZnO decreases after doping Cu [11]. Effect of the cu source on optical properties of CuZnO films deposited by ultrasonic spraying has been investigated. According to that study, CuZnO films synthesized from copper acetate solution exhibits better properties [12].

CuO exhibits Room Temperature Ferromagnetism [4], and thus this material offers a very good option for a class of spintronics even without the presence of any transition metal. The magnetic properties of thin films have been theoretically explained using Heisenberg Hamiltonian by us [13-17]. Thin films of $Cu_2O$/CuO [18, 19], ZnO [20] and carbon nanotubes [21] have been synthesized by us using expensive technique incorporated with vacuum. In our previous deposition techniques, pure Cu target was used to deposit CuO or $Cu_2O$ films by means of reactive dc sputtering. Amount of oxygen left in the vacuum of sputter coater was sufficient to crystallize the phase of CuO or $Cu_2O$. However, spin coating technique is found be low cost, reliable and fast. Thin films of Zn doped CuO films described in this manuscript were fabricated using spin coating technique. The band gap described in this manuscript was calculated using an optical method. However, the band gap and activation energy can be found by measuring thermal conductivity of sample at different temperatures for doped nanoparticles according to our previous studies [22].

**2. Experimental:**

Diethanolamine, cupric acetate and isopropyl alcohol were initially stirred together. The mole ratio between cupric acetate and diethanolamine was 1:1. Then a solution with $Cu^{+2}$ concentration of 1.5 mol dm$^{-3}$ was prepared with isopropyl alcohol as the solvent. Zinc acetate was added to the solution as the $Zn^{2+}$ agent. Different weight percentages of $Zn^{2+}$ from 4% to 10% in steps of 2% were added to the solutions with $Cu^{2+}$. Then these solutions were continuously stirred for 24 hours. This stirred solution was applied on glass substrates for sol-gel spin coating. Films were synthesized at three different spin speeds (1500, 2200 and 2400 rpm) for two different time durations (15 and 60 sec.). Spin coated thin films were subsequently annealed in air at temperatures of 150, 250, 350, 450 and 500 °C for 1 and 2 hours. Films were deposited on normal non-conducting glass substrates were employed for structural and optical characterization. Some films were fabricated on conducting glass substrates for electrical characterization.

Optical properties of films were determined using Shimadzu 1800 UV-visible spectrometer. Structural properties were investigated using X-ray diffractometer Rigaku Ultima IV with Cu-K$_\alpha$ radiation.

**3. Results and discussion:**

Films fabricated at spin speed of 2200rpm for 1 min were found to be uniform. After annealing samples at 500 $^0$C for one hour in air, the phase of CuO could be crystallized. Therefore, all the films explained in this manuscript were synthesized at 2200 rpm for 1 min and subsequently annealed at 500 $^0$C for one hour. X-ray diffraction (XRD) method was applied to determine the structural properties of samples. Figure 1 shows the XRD patterns of pure CuO samples and CuO samples doped with 4% and 10% of $Zn^{2+}$.

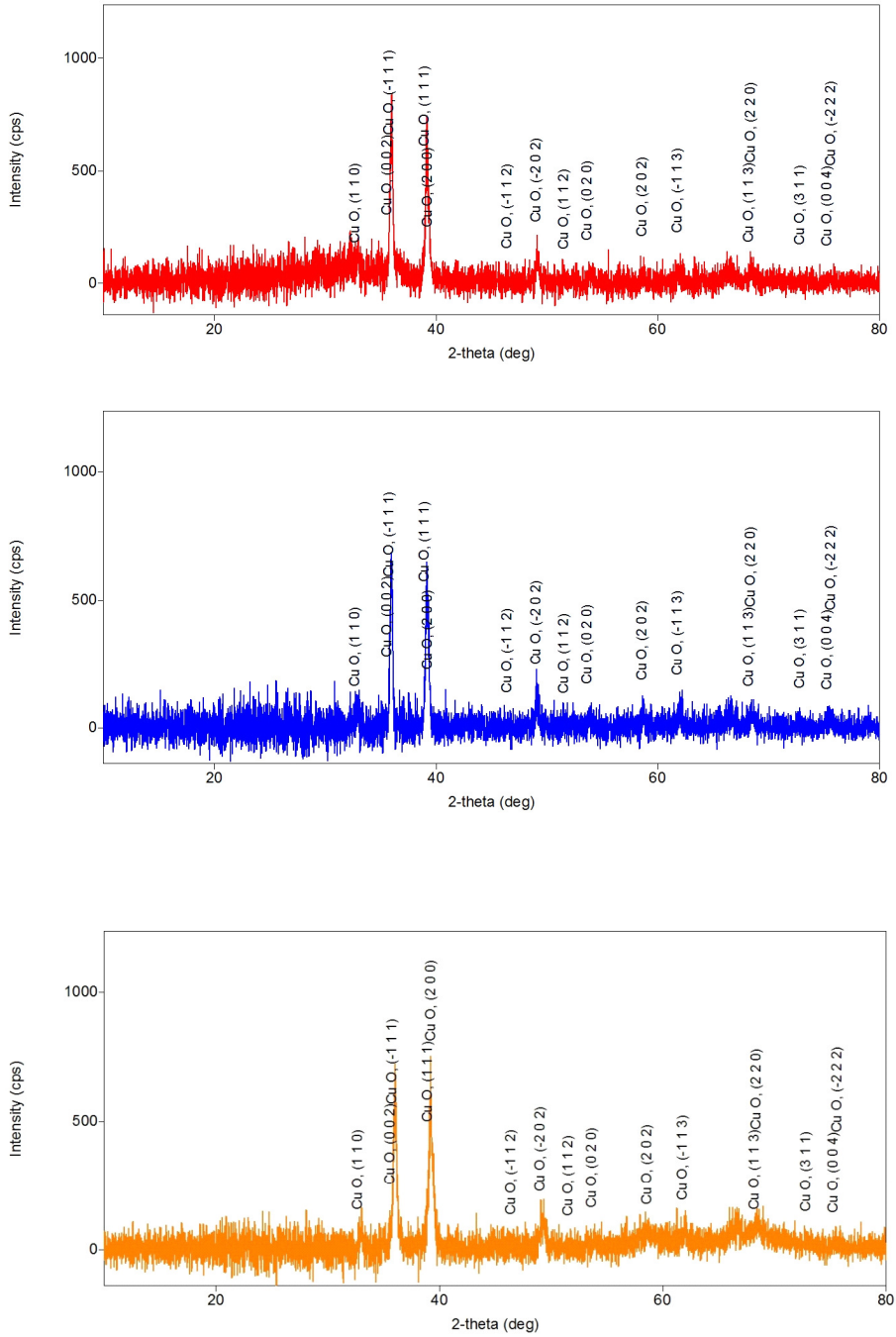

Figure 1: XRD patterns of pure CuO- top, 4%Zn-center and 10%Zn-bottom

Table 1 shows the particle size, dislocation density and strain determined using (111) peak of XRD patterns given in figure 1. Particle size (D) was calculated using

$$D = \frac{0.91\lambda}{\beta \cos\theta} \quad (1)$$

Where $\lambda$ is the wavelength of Cu-K$_\alpha$ radiation ($\lambda$=1.54060 $^0$A) and $\beta$ is the full width at half maximum (FWHM) of XRD peak at angle $\theta$.

The dislocation density ($\delta$) was determined using

$$\delta = \frac{1}{D^2} \quad (2)$$

The strain ($\varepsilon$) was calculated using

$$\varepsilon = \frac{\beta \cos\theta}{4} \quad (3)$$

| Zn$^{2+}$ doping concentration | angle $\theta$ (deg) | FWHM (deg) | Particle size (nm) | Dislocation density ($10^{16}$ lines/m$^2$) | Strain |
|---|---|---|---|---|---|
| 0% | 17.985 | 0.250 | 8.641 | 0.0134 | 0.0406 |
| 4% | 17.984 | 0.237 | 9.126 | 0.0120 | 0.0384 |
| 6% | 17.977 | 0.234 | 9.319 | 0.0115 | 0.0376 |
| 10% | 17.990 | 0.327 | 6.569 | 0.0232 | 0.0534 |

Table 1: Variation of particle size, dislocation density and strain with doping concentration.

Figure 2 shows the variation of particle size with doped Zn$^{2+}$ percentage. The particle size increases up to nearly 6%, and then decreases. Dislocation density and strain decrease with the increase of particle size, and they increase with the decrease of particle size. Adding an atom with larger radius can increase the particle size up to some concentration. Adding so many doping atoms can break the larger particles, and thereafter particle size will decrease. Strain is the relative change of particle size ($\frac{\Delta D}{D}$). Because the change of particle size ($\Delta D$) doesn't vary considerably, the strain ($\frac{\Delta D}{D}$) decreases with the increase of particle size ($D$). Because dislocations and deformations (stress in this case) correlate to each other, they vary in the same manner.

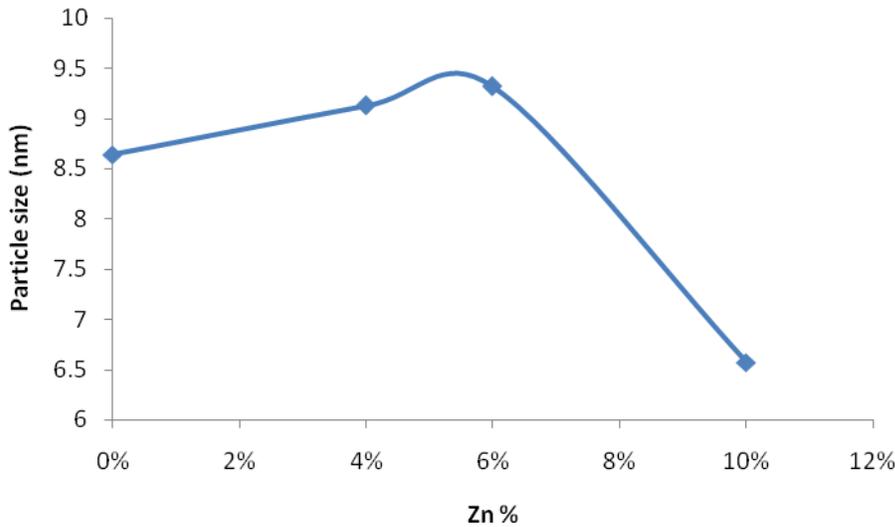

Figure 2: Particle size versus doped $Zn^{2+}$ percentage

The relative intensity of (111) peak in XRD patterns gradually increases with the doping concentration. Only the phase of CuO can be found in all the XRD patterns. Possible reasons are as follows. When $Zn^{2+}$ occupies interstitial sites or vacant sites in CuO lattice, peaks of $Zn^{2+}$ can't be found in XRD patterns. Only a variation of relative intensities of existing peaks can be observed in that case. The phase of Zn can be amorphous in thin film. Even if Zn is crystallized in the thin film, it can be crystallized at grain boundaries. The weak XRD peaks of trace amount of Zn are not noticeable.

UV/Vis absorption spectrums are shown in figure 3 at different $Zn^{2+}$ doping concentrations. Absorption edge varies from 869 to 949 nm according to these spectrums as $Zn^{2+}$ doping concentration is varied from 0 to 10%. Tauc plots are given in figure 4 for different doping concentrations. Optical band gap was determined using the intercept of tangential line on horizontal axis as shown in the figure 4. Here the dashed line represents the tangential line drawn to the curve at the absorption edge. Table 2 shows the variation of optical band gap with doping concentration.

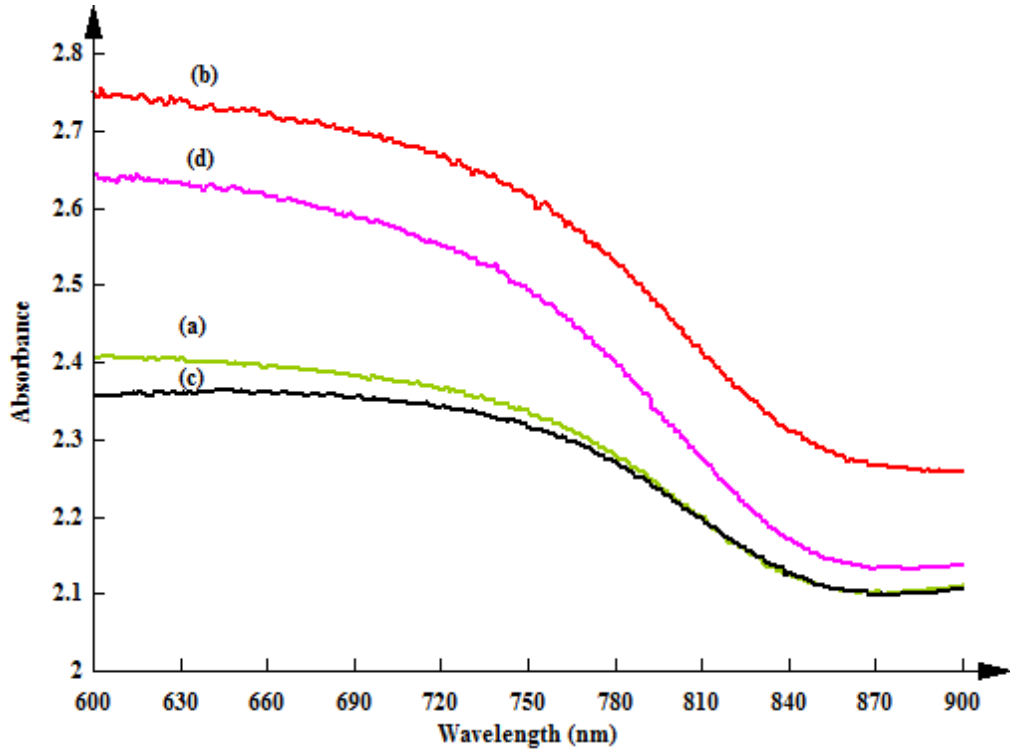

Figure 3: UV-Vis absorbance for (a) Pure Cu, (b) Zn 10% Cu 90%, (c) Zn 6% Cu 94% and (d) Zn 4% Cu 96%

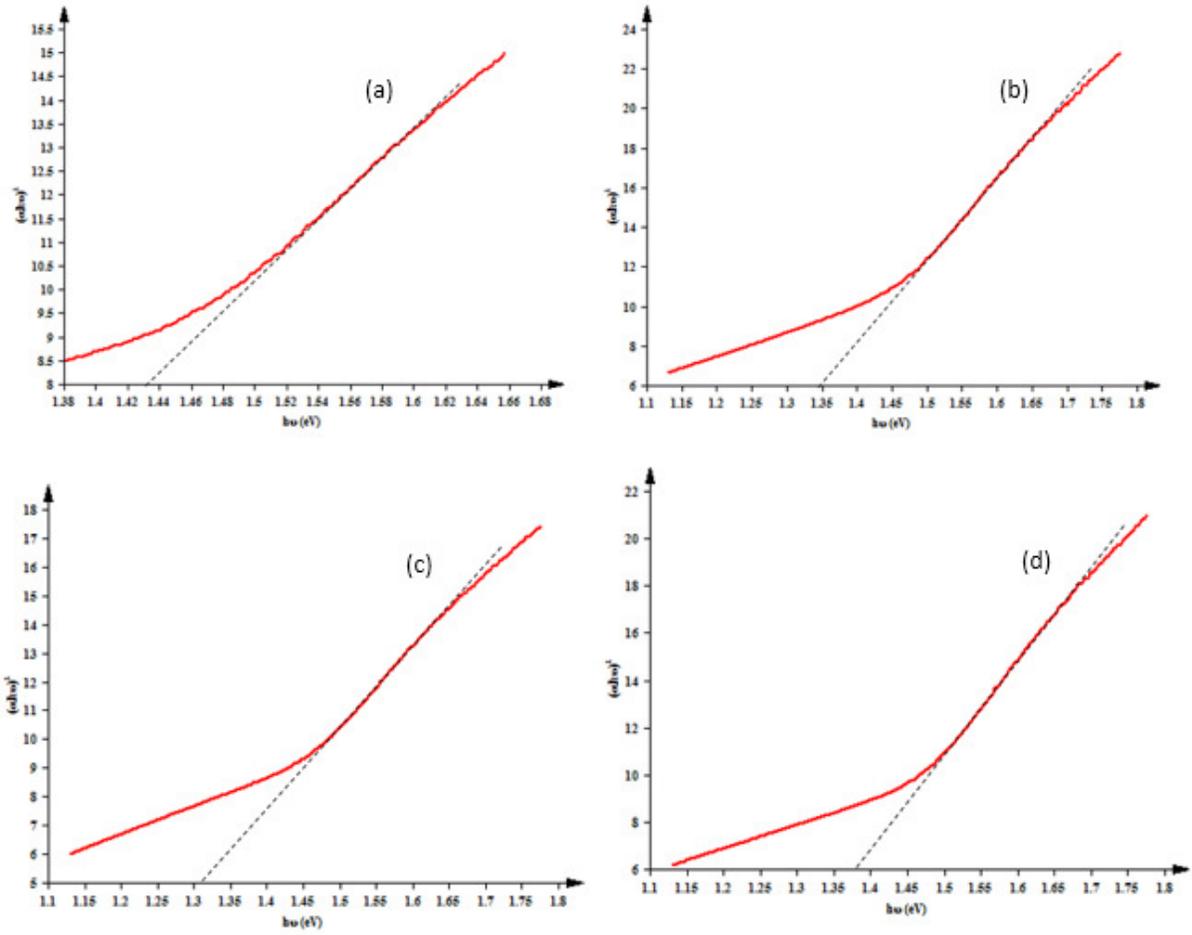

Figure 4: (αhν)² versus (hν) plots for (a) Pure Cu, (b) Zn 10% Cu 90%, (c) Zn 6% Cu 94% (d) Zn 4% Cu 96%

| Doping concentration | Optical band gap (eV) |
| --- | --- |
| (a) Pure Cu | 1.43 |
| (b) Zn 10% Cu 90% | 1.34 |
| (c) Zn 6% Cu 94% | 1.31 |
| (d) Zn 4% Cu 96% | 1.38 |

Table 2: Variation of optical band gap with doping concentration

Optical band gap decreases with doping concentration up to nearly 6% as expected. Because impurity atoms add some energy levels to band structure, the optical band gap decreases. The scattering of conduction electrons will increase as the field inside the crystal is fluctuated due to the existence of high concentration of impurity atoms. Therefore, the optical band gap increases above doping concentration of nearly 6%. Figure 5 shows the variation of optical band gap with $Zn^{2+}$ doping concentration.

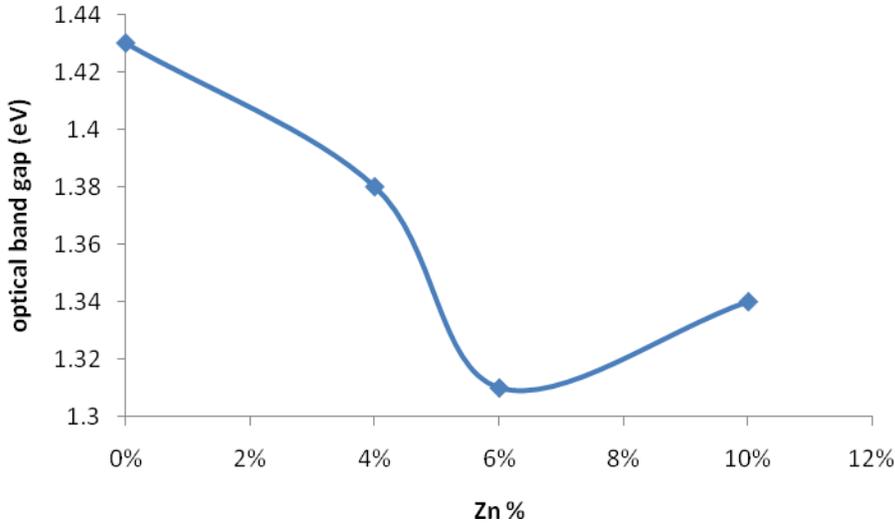

Figure 5: Optical band gap versus $Zn^{2+}$ doping concentration

According to figure 2 and figure 5, the optical band gap decreases with the increase of particle size. The relationship between particle radius and the band gap is given by Brus equation as follows.

$$E_g = E_{bulk} + \frac{h^2}{8r^2}(1/m^*_e + 1/m^*_h) - \frac{1.786e^2}{4\pi\varepsilon_0\varepsilon_r r^2} \qquad (4)$$

Where $E_g$ and $E_{bulk}$ are the band gaps thin film and bulk, respectively, h is the Planck's constant, $m^*_e$ and $m^*_h$ are the effective masses of electron and hole, respectively, $\varepsilon$ is the dielectric constant and r is the radius of the particle. Therefore, the optical band gap is inversely proportional to particle size.

## 4. Conclusion:

According to XRD patterns, all the film samples contain the single phase of CuO. Because doped $Zn^{2+}$ ions occupy the interstitial or vacant sites in the cell of CuO, Zn doesn't contribute to XRD peaks. The particle size increased up to the doping concentration of 6% approximately, and then decreased. Both dislocation density and strain decreased and then increased with doping concentration. The optical band gap first decreased and then increased with doping concentration. This behavior was explained using Brus equation. The band gap of CuO can be reduced by adding a trace amount of Zn, and materials with lower band gap find potential applications in photovoltaic applications. As explained in the introduction part, the band gap of CuO is in the range of 1.2-1.9 eV. The optical band gap values of our $Zn^{2+}$ doped CuO samples also lie within this range. However, the doping of higher amount of $Zn^{2+}$ increased the band gap. The increase of scattering of conduction electrons at higher doping concentration is the reason for this behavior of optical band gap.

**References:**

1. Guanhua Chen, Jean-Marc Langlois, Yuejin Guo, and William A. Goddard, III, 1989. Superconducting properties of copper oxide high temperature superconductors. Proceedings of the National Academy of Sciences of U S A 86 (10), 3447–3451.
2. D. Arun Kumar, Francis P. Xavier and J. Merline Shyla, 2012. Investigation on the variation of conductivity and photoconductivity of CuO thin films as a function of layers of coating. Archives of Applied Science Research 4 (5), 2174-2183.
3. Jeong-Woo Park, Kang-Jun Baeg, Jieun Ghim, Seok-Ju Kang, Jeong-Ho Park and Dong-Yu Kim, 2007. Effects of copper oxide/gold electrode as the source-drain electrodes in organic thin film transistors. Electrochemical and Solid-State Letters 10 (11), 340-343.
4. Daqiang Gao, Jing Zhang, Jingyi Zhu, Jing Qi, Zhaohui Zhang, Wenbo Sui, Huigang Shi and Desheng Xue, 2010. Vacancy-mediated magnetism in pure copper oxide nanoparticles. Nanoscale Research Letters 5, 769–772.
5. D.S.C. Halin, I.A. Talib, M. A. A. Hamid and A.R. Daud, 2008. Characterization of cuprous oxide thin films on n-Si substrate prepared by sol-gel spin coating. ECS Journal of Solid State Science and Technology 16 (1), 232-237.
6. T. Mahalingam, V. Dhanasekaran, G. Ravi, Soonil Lee, J. P. Chu and Han-Jo Lim, 2010. Effect of deposition potential on the physical properties of electrodeposited CuO thin films. Journal of optoelectronics and Advanced Materials 12 (6), 1327-1332.
7. V. Saravanakannan and T. Radhakrishnan, 2014. Structural, electrical and optical characterization of CuO thin films prepared by spray pyrolysis technique. International Journal of ChemTech Research CODEN( USA) 6 (1), 306-310.
8. J.S. Bhat, K.I. Maddani and A.M. Karguppikar, 2006. Influence of Zn doping on electrical and optical properties of multilayered tin oxide thin films. Bulletin of Materials Science 29(3), 331-337.
9. Mohamed A. Basyooni, Mohamed Shaban and Adel M. El Sayed, 2017. Enhanced gas sensing properties of sip coated Na doped ZnO nanostructured films. Nature, doi:10.1038/srep41716.
10. Ibrahim Y. Erdogan, 2010. The alloying effects on the structural and optical properties of nanocrystalline copper zinc oxide thin films fabricated by spin coating and annealing method. Journal of Alloys and Compounds 502(2), 445-450.

11. Shaveta Thakur, Neha Sharma, Anamika Varkia and Jitender Kumar, 2014. Structural and optical properties of copper doped ZnO nanoparticles and thin films. Advances in Applied Science Research 5(4), 18-24.
12. Chih-Hung Hsu, Lung-Chien Chen and Xiuyu Zhang, 2014. Effect of the cu source on optical properties of CuZnO films deposited by ultrasonic spraying. Materials 7, 1261-1270.
13. P. Samarasekara and Udara Saparamadu, 2012. Investigation of Spin Reorientation in Nickel Ferrite Films. Georgian electronic scientific journals: Physics 1(7), 15-20.
14. P. Samarasekara and N.H.P.M. Gunawardhane, 2011. Explanation of easy axis orientation of ferromagnetic films using Heisenberg Hamiltonian. Georgian electronic scientific journals: Physics 2(6), 62-69.
15. P. Samarasekara, 2008. Influence of third order perturbation on Heisenberg Hamiltonian of thick ferromagnetic films. Electronic Journal of Theoretical Physics 5(17), 227-236.
16. P. Samarasekara and Udara Saparamadu, 2013. In plane oriented Strontium ferrite thin films described by spin reorientation. Research & Reviews: Journal of Physics-STM journals 2(2), 12-16.
17. P. Samarasekara and B.I. Warnakulasooriya, 2016. Five layered fcc ferromagnetic films as described by modified second order perturbed Heisenberg Hamiltonian. Journal of science: University of Kelaniya 11, 11-21.
18. P. Samarasekara, 2010. Characterization of Low Cost p-$Cu_2O$/n-CuO Junction. Georgian Electronic Scientific Journals: Physics 2(4), 3-8.
19. P. Samarasekara and N.U.S. Yapa, 2007. Effect of sputtering conditions on the gas sensitivity of Copper Oxide thin films. Sri Lankan Journal of Physics 8, 21-27.
20. P. Samarasekara, A.G.K. Nisantha and A.S. Disanayake, 2002. High Photo- Voltage Zinc Oxide Thin Films Deposited by DC Sputtering. Chinese Journal of Physics 40(2), 196-199.
21. P. Samarasekara, 2009. Hydrogen and Methane Gas Sensors Synthesis of Multi-Walled Carbon Nanotubes. Chinese Journal of Physics 47(3), 361-369.
22. K. Tennakone, S.W.M.S. Wickramanayake, P. Samarasekara and, C.A.N. Fernando, 1987. Doping of Semiconductor Particles with Salts. Physica Status Solidi (a)104, K57-K60.
12